\documentclass{ws-procs9x6}

\begin{document}

\title{Effective string picture for confining gauge theories
at finite temperature
}

\author{M.~CASELLE and M.~PANERO
\footnote{\uppercase{P}resenter of the talk.}
}

\address{Universit\`a di Torino and INFN, Sezione di Torino,\\
Via P. Giuria, 1\\
I--10125 Torino, Italy\\
E-mail: caselle@to.infn.it, panero@to.infn.it
}

\author{M.~HASENBUSCH}

\address{NIC/DESY Zeuthen\\
Platanenallee 6,\\
D--15738 Zeuthen, Germany\\
E-mail: Martin.Hasenbusch@desy.de
}

%%%%%%%%%%%%%%%%%%%%%%%%%%%%%%%%%%%%%%%%%%%%%%%%%%%%%%%%%%%%%%
% You may repeat \author \address as often as necessary      %
%%%%%%%%%%%%%%%%%%%%%%%%%%%%%%%%%%%%%%%%%%%%%%%%%%%%%%%%%%%%%%

\maketitle

\abstracts{We study the ``effective string picture'' of
confinement, deriving theoretical predictions
for the interquark potential 
at finite temperature.
At low temperatures,
the leading string
correction to the linear confining potential
between a heavy quark-antiquark pair is
the ``L\"uscher term''. Assuming
a Nambu--Goto
effective string action, subleading contributions
can be worked out in an analytical way. We also discuss the
contribution given by a possible ``boundary term'' in the
effective action, and compare these predictions with high
precision results from simulations of lattice $\mathbf{Z}_2$ gauge
theory in three dimensions, obtained with an algorithm that
exploits the duality of the $\mathbf{Z}_2$ gauge model with the
Ising spin model.}

\section{Effective String Picture of Confinement}
%The basic idea behind the effective string framework for
%confinement is the following: two color charges
%in the confined regime of a gauge theory behave as if they
%were joined by a thin flux tube, which can fluctuate like a
%vibrating string.
%Such a flux tube is considered as a
%one-dimensional object, and it is associated with a given
%string tension, which is responsible for the confining force
%which ties the quark--antiquark pair.
The string
picture is an effective framework
%(with no \emph{direct} connection with the
%fundamental degrees of freedom)
which is expected to provide a good physical description of the
infrared behavior of confining gauge theories. The basic idea is
simple: two confined color charges behave as if they were joined
by a thin flux tube, which can fluctuate like a vibrating string.
The dynamics of the world sheet spanned by the string during its
time-like evolution is described by an effective action
$S_{\mbox{\tiny{eff}}}$, and for \emph{massless} string
fluctuations, the simplest choice for a candidate effective action
is the Nambu--Goto string action: $ S_{\mbox{\tiny{eff}}} = \sigma
\cdot \mathcal{A} $,
%\begin{equation}
%S_{\mbox{\tiny{eff}}} = \sigma \cdot \mathcal{A}
%\end{equation}
which is proportional to the area $\mathcal{A}$ of the
world--sheet surface; $\sigma$ is the \emph{string tension},
%% MH "a parameter" to "the parameter"
appearing as the parameter of the effective theory.
%;
%its value depends on the value of the fundamental coupling in
%the underlying gauge theory.

For a three-dimensional system with extension $L_s^2 \times L$
(with $L_s \gg L$) and periodic boundary conditions along the short
direction, the temperature $T$ is proportional to $1/L$, and in
that case
%one can integrate over the possible surface
%configurations of the string world sheet joining two
%Ployakov loops, by keeping the leading terms in the world
%sheet area, and
the result for the
expectation value of the Polyakov loop
correlation function reads:
\begin{equation}
\langle P^\dagger (R) P(0) \rangle =
\frac{ e^{-\sigma RL + k} }{\eta \left( i\frac{L}{2R} \right)}
\end{equation}
where $\eta$ is Dedekind's function.
%\begin{equation}
%\eta(\tau) = q^{\frac{1}{24}} \prod_{n=1}^{+\infty} (1-q^n)
%\;\;\;\;    ;      \;\;\;\;  q=e^{2\pi i \tau}
%\end{equation}
The term associated with the minimal world sheet
surface induces the
exponential area-law falloff responsible for
the linear rise in the interquark potential $V(R)$, while
%the (regularized) functional integration of
the first non-trivial contribution in $S_{\mbox{\tiny{eff}}}$
results in the determinant of the Laplace operator, and the
corresponding contribution to the interquark
potential $V(R)$ --- in a regime of distances shorter
than $\frac{L}{2}$ ---
is the L\"uscher term\cite{lsw}:
\begin{equation}
V(R)=-\frac{1}{L} \ln \langle P^\dagger (R) P(0) \rangle \simeq \sigma R -\frac{\pi}{24 R}
\end{equation}

%This result can be generalized in various ways: for
%instance, one may include
Inclusion of further terms\cite{noi} in the expansion
of the world sheet area results\footnote{\uppercase{T}his calculation involves a
%% MARCO OCT 7TH:
%% ERASEN "Riemann's"
%Riemann's 
$\zeta$ function regularization.} in a
contribution involving a combination of Eisenstein
functions\cite{serre}:
\begin{equation}
\label{nlo} -\frac{\pi^2}{1152 \sigma R^3} \left[2E_4 \left(
i\frac{L}{2R} \right) -E_2^2 \left( i\frac{L}{2R} \right)\right]
\end{equation}
However, such a contribution is still under 
%debate.
debate\cite{noi}.

On the other hand, it is also possible to include
a ``boundary term\footnote{\uppercase{S}uch
a boundary term is related to
derivatives of the $h$ field (which describes
transverse displacements with respect to the minimal
area surface of the world sheet), evaluated along
the Polyakov lines}'' in the
effective action:
%\begin{equation}
%S_{\mbox{\tiny{eff}}}= \sigma \cdot \left ( \mathcal{A} +
%\frac{b}{4} \int_0^L dt [ (\partial_z h)_{z=0}^2 +
%(\partial_z h)_{z=R}^2 ] \right)
%\end{equation}
%this results in a contribution to the behavior of the
%interquark potential in the ``low temperature regime''
%reading: $-\frac{b \pi }{24 R^2}$, where $b$ is
%related to the coefficient of the ``boundary term''
%in the effective action. This result can be obtained
%by
a perturbative expansion in $b$ (a parameter proportional
to the coefficient of the ``boundary term''
in the effective action), induces a leading order correction
like:
\begin{equation}
R \longrightarrow \frac{R}{\sqrt{1+ \frac{2b}{R}}}
\end{equation}
with a short distance contribution to $V(R)$ reading: $-\frac{b
\pi }{24 R^2}$.
\section{The Model: $\mathbf{Z}_2$ Lattice Gauge Theory}
%To test the predictions from the effective string
%picture,
We run numerical simulations of the $\mathbf{Z}_2$ lattice
gauge theory in three space-time dimensions.
%a very simple,
%though non-trivial confining model.
This choice
%of a discrete gauge group model
has various
motivations: the effective string picture
is believed to be independent of the
underlying gauge group; the $\mathbf{Z}_2$
gauge group is interesting from the perspective of
the \emph{center} role in confinement\footnote{$\mathbf{Z}_2$
is the center of continuous gauge groups like $SU(2)$ or
$Sp(N)$.}; the reduced configuration space
%of this simple
%theory
of this theory and its duality with respect to the Ising spin
model enable one to get high precision results within a reasonable
amount of CPU time.

The pure 3D lattice gauge model is described in terms of
$\sigma_{x,\mu}$ variables (taking values in $\mathbf{Z}_2$)
defined on the lattice bonds; the dynamics is governed by the
standard Wilson action, which enjoys $\mathbf{Z}_2$ gauge
invariance\footnote{$\mathbf{Z}_2$ gauge transformations act as
\emph{local} flips of $\sigma_{x,\mu}$ variables living on the
lattice bonds which meet at a given site.}. The partition function
reads:
\begin{equation}
Z(\beta)= \sum_{\mbox{c.} } e^{ -\beta S }
= \sum_{\mbox{c.} }
\exp \left[ +\beta \sum_{\Box} \sigma_{\Box} \right]
\end{equation}
and the system may exist in different phases: a confined, strong
coupling phase, with massive string fluctuations for $\beta <
0.47542(1)$ \cite{HaPi}; a confined, \emph{rough} phase, with
massless string fluctuations (this is the regime we studied in our
simulations); a deconfined phase for $\beta > 0.7614134(2)$
\cite{BlTa}.

This model is \emph{dual} with respect to the $\mathbf{Z}_2$ spin
model in 3D, and
%; the mapping reads:
%\begin{equation}
%Z_{\mbox{\tiny{gauge}}}(\beta) \propto Z_{\mbox{\tiny{spin}}}(\tilde\beta)
%\;\; , \;\;\;\; \mbox{with } \; \tilde{\beta}=-\frac{1}{2} \ln \tanh \beta
%\end{equation}
%and VEV's of
%Polyakov loop correlators in the gauge model can be written
%as ratios between partition functions of the
%spin system
%:
%\begin{equation}
%G(R)= \langle P^\dagger (R) P(0) \rangle =
%\frac{ Z^{(S)}_{\mbox{\tiny{spin}}} (\tilde{\beta}) }{ Z_{\mbox{\tiny{spin}}} (\tilde{\beta})}
%\end{equation}
%where $Z^{(S)}_{\mbox{\tiny{spin}}}(\tilde{\beta})$ is
%the partition function of the spin system
%with stacks
%of ``antiferromagnetic couplings''.
%the \emph{snake
%algorithm}
we exploited this property to express a ratio between Polyakov
%%MH $G(R)=\langle P^\dagger (R) P(0) \rangle$ added to define G(R)
%% still most things in the eqaution below are undefined.
loop correlators $G(R)=\langle P^\dagger (R) P(0) \rangle$ 
of the gauge theory as a product of expectation
values of one-link variables in the modified spin ensembles. 
%% MARCO SEPT. 25TH:  
%% SKIPPING EQUATION WHICH EXPRESSES 
%% CORRELATOR RATIOS AS A PRODUCT OF VEV'S 
%values of one-link variables in the modified spin ensembles:
%\begin{equation}
%\frac{G(R)}{G(R+1)} = \frac{Z_{L \times R } }{Z_{L \times (R+1) }}
%= \frac{Z_{L \times R } }{Z_{[L \times R] + 1 }} \dots
%\frac{Z_{ [L \times R] + m} }{Z_{[L \times R] + (m+1) }} \dots
%\frac{Z_{ [L \times R] + L-1} }{Z_{L \times (R+1) }}
%\end{equation}
A similar algorithm is the so-called \emph{snake
algorithm}\cite{snake}.

We used multi-level updating
% to measure the ratios,
and a hierarchical organization of sublattices, and the CPU time
turns out to be roughly proportional to the inverse temperature
$L$, and \emph{virtually independent of the distance $R$ between
the quark sources}, thus the algorithm is particularly
useful in a regime of very large interquark distances.
%Moreover,
%there is no cross-correlation among output results.

\section{Numerical Results}
Let $F(R,L)$ be the free energy associated with the
presence of a heavy quark-antiquark pair at finite
temperature: $ G(R)=e^{-F(R,L)} $. We studied
``quantum contributions''
in free energy differences, by measuring
the following quantity:
\begin{equation}
Q(R,L) = F(R+1,L) - F(R,L) - \sigma L
\end{equation}

%In the ``large interquark
%distances'' regime ($L<2R$), we found good agreement with
%the predictions from the Nambu--Goto effective
%action truncated at the next-to-leading order (although
%in principle a next-to-leading order renormalization of
%the string tension $\sigma$ should be discussed).
%We also found that the coefficient for a possible ``boundary
%term'' in the effective action for this model seems to be
%very small, likely zero.
\begin{figure}[ht]
%\epsfxsize=10cm   %width of figure - will enlarge/reduce the figures
%\epsfbox{fig3.eps}
%\figurebox{2cm}{3cm}{} %to have a box alone
\centerline{\epsfxsize=3.9in\epsfbox{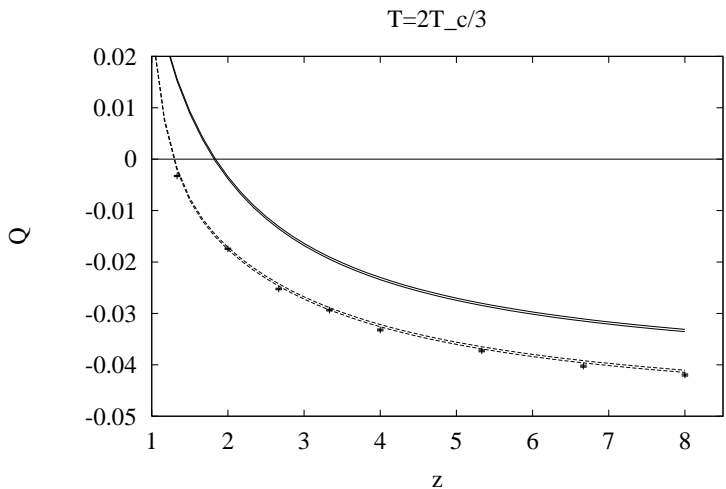}}
\caption{$Q(R,L)$ for $L=12$ (i.e. $T=2T_c/3$)
at $\beta=0.75180$. $z$ is defined as:
$z\equiv \frac{2R}{L}$. Solid curves correspond to the
free bosonic string prediction,
while dashed lines
%% MARCO OCT 7TH:
%% REPLACED "correspond to" WITH "include"
%correspond to 
include 
the LO Nambu--Goto correction. Pure
area--law corresponds to $Q=0$. \label{23tc} }
\end{figure}
%\begin{figure}[ht]
%\epsfxsize=10cm   %width of figure - will enlarge/reduce the figures
%\epsfbox{fig3.eps}
%\figurebox{2cm}{3cm}{} %to have a box alone
%\centerline{\epsfxsize=3.9in\epsfbox{fig4.eps}}
%\caption{$Q(R,L)$ for $L=10$ (i.e. $T=4T_c/5$)
%at $\beta=0.75180$. \label{45tc} }
%\end{figure}

Fig. \ref{23tc} shows that at ``high temperatures''
($L<2R$)
%a comparison between the numerical
%results (with their errorbars), the LO-truncated
%(solid lines) and NLO-truncated Nambu--Goto string
%predictions; in this regime,
our numerical results are in
good agreement with the NLO prediction from
the Nambu--Goto string, while a
pure area law
%(which would correspond to $Q=0$)
is definitely ruled out, and the LO term alone
is not sufficient to describe the data.
%These results become
%even more impressive when one considers temperatures
%closer to the critical deconfinement
%temperature.
We also found that the coefficient of a possible ``boundary term''
%in the effective action
for this model seems to be
%% MH : most probably to likely.
very small, likely  zero.

%It is
%important to stress that we did not make any fit
%:
%the theoretical curves are plotted using the string tension
%value which is known from the literature, and the (small)
%width defined by the theoretical curves takes into account
%the uncertainty in the value of $\sigma$.
%
%In the ``short distance'' regime ($2R<L$),
%we found very good agreement between numerical results
%and the theoretical prediction of the Nambu--Goto string
%truncated at leading order --- which in this regime is
%the L\"uscher term.
\begin{figure}[ht]
%\epsfxsize=10cm   %width of figure - will enlarge/reduce the figures
%\epsfbox{fig3.eps}
%\figurebox{2cm}{3cm}{} %to have a box alone
\centerline{\epsfxsize=3.9in\epsfbox{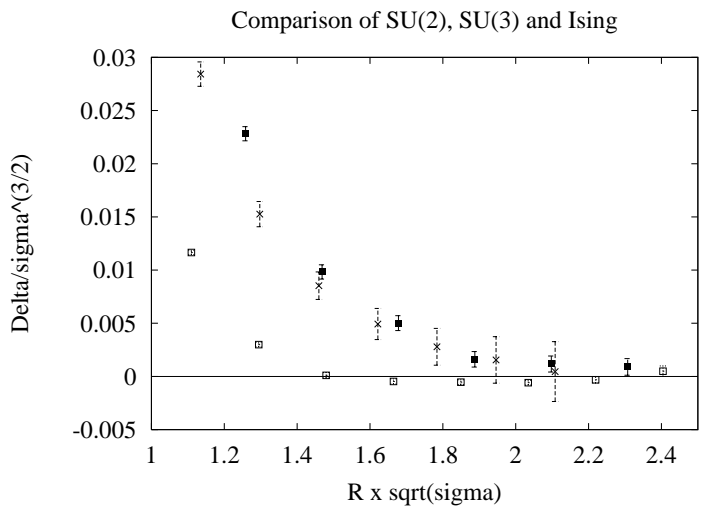}}
\caption{Behavior of different gauge models in the 
%% MARCO OCT 7TH:
%% REPLACED "short distance" WITH "low temperature"
%short distance
low temperature
regime. \label{newdelta_zoom} }
\end{figure}
As it concerns the $(L>2R)$ regime,
Fig. \ref{newdelta_zoom} shows the deviation\cite{noi} of a
quantity proportional to $\left[Q(R,L)-Q(R-1,L)\right]$ from the free
string prediction. The chosen normalization 
allows a
meaningful comparison among different LGT's in 3d: $SU(2)$ gauge
theory\cite{su2} (crosses), $SU(3)$ gauge 
%% MARCO OCT 7TH:
%% REPLACED "model" WITH "theory"
%model\cite{su3}
theory\cite{su3}
(white squares), and $\mathbf{Z}_2$ gauge theory\cite{noi} 
% MARCO SEPT. 25TH: ADDED SOME COMMENTS ABOUT THE LOW TEMPERATURE DATA FIGURE.
(black squares). The three models display the same qualitative behavior,
and, in particular, the data for $\mathbf{Z}_2$ and $SU(2)$
(which are groups with the same center --- namely: $\mathbf{Z}_2$ itself)
are compatible within errorbars. This may be a signature of the relevance of
center degrees of freedom to the confinement mechanism.
%The agreement between
%from the onset of the effective string, the
%deviation from the LO prediction of the Nambu--Goto action is
%compatible with zero.
%our $\mathbf{Z}_2$ results, show fairly constant errorbars,
%independent of $R$.
%Moreover,
%$\mathbf{Z}_2$ and $SU(2)$ data can be a possible signature of the
%center relevance to confinement.

\section{Conclusions}
We studied confining gauge theories at finite temperature, and
tested the theoretical predictions
%that can be
%formulated within the framework of an
of the Nambu--Goto effective string for $\mathbf{Z}_2$ lattice
gauge theory, both at large and short interquark distances. Our
algorithm exploits the duality of the model, and this enabled us
to explore a wide range of distances, detecting next-to-leading
order effects. Our data seem to rule out a ``boundary term'' in
the effective string action describing the present gauge model.
Finally, we also made a comparison with some different gauge
models.

%%%%%%%%%%%%%%%%%%%%%%%%%%%%%%%%%%%%%%%%%%%%%%%%%%%%%%%%%%%%%%%%%%%%%%%
%
%Use this if your figures are put in a subdirectory having the same
%name as the main latex file, ie:
%
%      ws-procs9x6/procs-fig1.eps
%      ws-procs9x6/procs-fig2.eps
%      ws-procs9x6/procs-fig3.eps
%      ws-procs9x6/procs-fig4.eps
%      etc.
%
%\begin{figure}[htbp] %ORIGINAL SIZE: width=1.4TRUEIN; height=1.5TRUEIN
%\figurebox{}{}{procf1} %100 percent
%\caption{Labeled tree {\it T}.}
%\end{figure}
%
%%%%%%%%%%%%%%%%%%%%%%%%%%%%%%%%%%%%%%%%%%%%%%%%%%%%%%%%%%%%%%%%%%%%%%%
%
%\begin{thebibliography}{0}
%\bibitem{lsw} M. L\"uscher, K. Symanzik and P. Weisz,
%{\it Nucl. Phys.} {\bf B173} (1980) 365.
%
%\bibitem{noi} See M. Caselle, M. Hasenbusch and M. Panero, {\it JHEP}
%{\bf 0301} (2003) 057, and references therein; see also M.
%Caselle, M. Panero and P. Provero, {\it JHEP} {\bf 0206} (2002)
%061, and M. Caselle, M. Hasenbusch and M. Panero, in preparation.
%
%\bibitem{HaPi}
%M. Hasenbusch and K. Pinn, {\it J.Phys.} {\bf A30} (1997) 63.
%
%\bibitem{BlTa}
%H.W.J. Bl\"ote, L.N. Shchur, A.L. Talapov, {\it Int. J. Mod.
%Phys.} {\bf C10} (1999) 1137.
%
%
%\bibitem{snake}  P. de Forcrand, M. D'Elia and M. Pepe,
%{\it Phys. Rev. Lett.} {\bf 86} (2001) 1438; P. de Forcrand, M.
%D'Elia and M. Pepe, {\it Nucl. Phys. Proc. Suppl.} {\bf 94} (2001)
%494.
%
%\bibitem{su2} M. Caselle, M. Pepe and A. Rago, in preparation.
%
%\bibitem{su3} M. L\"uscher and P. Weisz, {\it JHEP}
%{\bf 0207} (2002) 049.
%
%
%\end{thebibliography}


\begin{thebibliography}{0}

\bibitem{lsw} M. L\"uscher, K. Symanzik and P. Weisz,
{\it Nucl.Phys.} {\bf B173} (1980) 365.
%%CITATION = NUPHA,B173,365;%%

\bibitem{noi} M. Caselle, M. Hasenbusch and M. Panero, {\it JHEP}
{\bf 0301} (2003) 057.
%%CITATION = HEP-LAT 0211012;%%
M. Caselle, M. Panero and P. Provero, {\it JHEP} {\bf 0206}, 061 (2002) 
%%CITATION = HEP-LAT 0205008;%% 
and references therein. 
M. Caselle, M. Hasenbusch and M. Panero, in
preparation.
\bibitem{serre}
J.P. Serre, ``A course in Arithmetic'', Springer--Verlag, New York, 1980.

\bibitem{HaPi}
M. Hasenbusch and K. Pinn, {\it J.Phys.} {\bf A30} (1997) 63.
%%CITATION = COND-MAT 9605019;%%

\bibitem{BlTa}
H.W.J. Bl\"ote, L.N. Shchur, A.L. Talapov, {\it Int.J.Mod.Phys.}
{\bf C10} (1999) 1137.
%%CITATION = COND-MAT 9912005;%%

\bibitem{snake}  
Ph. de Forcrand, M. D'Elia and M. Pepe,
{\it Phys.Rev.Lett.} {\bf 86} (2001) 1438. 
%%CITATION = HEP-LAT 0007034;%%
Ph. de Forcrand, M. D'Elia and M. Pepe, 
{\it Nucl.Phys.Proc.Suppl.} {\bf 94} (2001) 494.
%%CITATION = HEP-LAT 0010072;%%


\bibitem{su2} M. Caselle, M. Pepe and A. Rago, in preparation.


\bibitem{su3} M. L\"uscher and P. Weisz, {\it JHEP}
{\bf 0207} (2002) 049.
%%CITATION = HEP-LAT 0207003;%%

\end{thebibliography}
\end{document}